\documentclass[final,1p,times]{elsarticle}
\usepackage{color}

\journal{Astroparticle Physics}

\def\lsim{\mathrel{  
        \raise0.3ex\hbox{$<$}\kern-0.75em{\lower0.65ex\hbox{$\sim$}}}}

\begin{document}
\begin{frontmatter}

\title{A comparison study of CORSIKA and COSMOS simulations for extensive air showers}

\author[label1,label2]{Soonyoung Roh} \ead{syroh@canopus.cnu.ac.kr}
\author[label1]{Jihee Kim} \ead{jhkim@canopus.cnu.ac.kr}
\author[label1]{Dongsu Ryu\corref{cor1}} \ead{ryu@canopus.cnu.ac.kr}
\author[label3]{Hyesung Kang} \ead{hskang@pusan.ac.kr}
\author[label4]{Katsuaki Kasahara} \ead{kasahara@icrr.u-tokyo.ac.jp}
\author[label4]{Eiji Kido} \ead{ekido@icrr.u-tokyo.ac.jp}
\author[label5]{Akimichi Taketa} \ead{taketa@eri.u-tokyo.ac.jp}

\address[label1]{Department of Astronomy and Space Science, Chungnam National University, Daejeon 305-764, South Korea}
\address[label2]{Department of Physics, Graduate School of Science, Nagoya University, Nagoya 464-8602, Japan}
\address[label3]{Department of Earth Sciences, Pusan National University, Pusan 609-735, South Korea}
\address[label4]{Institute for Cosmic Ray Research, University of Tokyo, Chiba 277-8582, Japan}
\address[label5]{Center for High Energy Geophysics Research, Earthquake Research Institute, University of Tokyo, Tokyo 113-0032, Japan}
\cortext[cor1]{Corresponding author.}

\begin{abstract}
Cosmic rays with energy exceeding $\sim 10^{18}$ eV are referred to as ultra-high energy cosmic rays (UHECRs). Monte Carlo codes for extensive air shower (EAS) simulate the development of EASs initiated by UHECRs in the Earth's atmosphere. Experiments to detect UHECRs utilize EAS simulations to estimate their energy, arrival direction, and composition. In this paper, we compare EAS simulations with two different codes, CORSIKA and COSMOS, presenting quantities including the longitudinal distribution of particles, depth of shower maximum, kinetic energy distribution of particle at the ground, and energy deposited to the air. We then discuss implications of our results to UHECR experiments.
\end{abstract}

\begin{keyword}
extensive air shower \sep Monte Carlo simulation \sep ultra-high energy cosmic rays
\end{keyword}

\end{frontmatter}

\section{Introduction}

The nature and origin of ultra-high energy cosmic rays (UHECRs) with energy above $\sim 10^{18}$ eV are outstanding problems of modern physics. Many studies have been performed to unravel the problems: where do UHECRs come from, what is the composition of UHECRs, and how are UHECRs accelerated to such extreme energies? UHECRs are believed to be the result of extremely powerful cosmic phenomena \cite{Hillas}; the most powerful astrophysical events, such as active galactic nuclei (AGNs) \cite{Biermann-Strittmatter}, gamma ray bursts (GRBs) \cite{Waxman}, and shock waves around clusters of galaxies \cite{Kang}, have been suggested as possible sources of UHECRs. Yet, the nature and origin of UHECRs remain unsolved (see \cite{Berezinsky-etal,Kotera-Olinto} for review).

Cosmic rays (CRs), which are electrically charged particles, do not travel in straight lines in space. Their trajectories are bent by intergalactic and interstellar magnetic fields that are known to exist between galaxies and between stars \cite{Das-etal,Takami-Sato}. For this reason, even though we may guess their arrival directions at the Earth, we do not know exactly where they come from. However, this problem may be resolved if enough events of UHECRs are observed. 

Directly detecting UHECRs at the top of the Earth's atmosphere is practically impossible owing to the rarity of UHECR events. On average, only a few particles hit a square kilometer of the atmosphere per century. Such low flux of UHECRs demands experiments covering huge areas to increase opportunities for their detection. In the last few decades, UHECRs have been observed by Akeno Giant Air Shower Array (AGASA) \cite{Teshima-etal}, High Resolution Fly's Eye Experiment (HiRes) \cite{Abbasi-etal-04}, Pierre Auger Observatory (AUGER) \cite{Abraham-etal}, Telescope Array (TA) \cite{Sagawa-etal} and so on. These experiments detect extensive air showers (EASs) created by UHECRs.

When UHECRs enter the Earth's atmosphere, they first collide with air molecules of oxygen or nitrogen; subsequently through complex interactions and cascades, EASs, which are made of up to hundreds of billions of secondary particles (See Table 1 for the case of $10^{19.5}$eV primary), are generated \cite{Rossi-Greisen,Knapp}. By detecting the photons produced by secondary particles and/or the secondary particles arriving at the ground, the properties of primary particles such as the energy, arrival direction, and composition are inferred.

To observe UHECRs, AGASA used a ground array of scintillation detectors, HiRes used fluorescence telescopes, AUGER uses a hybrid facility of water Cherenkov tanks and fluorescence telescopes, and TA also uses a hybrid facility of scintillation detectors and fluorescence telescopes. Fluorescence telescopes measure the ultra-violet (UV) fluorescence light produced through interactions between air molecules and secondary particles in EASs, and arrays of scintillation detectors and water Cherenkov tanks recode the secondary particles arriving at the ground. From these, AGASA reported 58 events above 40 EeV \cite{Shinozaki-etal}. HiRes observed 13 events above 56 EeV \cite{Abbasi-etal-10}. AUGER and TA have so far reported 69 events \cite{Abreu-etal} and 15 events \cite{Tinyakov-etal} above 55 EeV and 57 EeV, respectively.

Along with observations of UHECRs, EASs need to be investigated by performing Monte Carlo (MC) simulations. EAS simulations form an essential part of UHECR experiments. The TA experiment employs two existing MC codes for the simulations, CORSIKA (COsmic Ray SImulations for KAscade) \cite{Heck-etal} and COSMOS \cite{Cohen-Kasahara, Kido-etal}. In this paper, we report a comparison study of CORSIKA and COSMOS simulations for TA, presenting the longitudinal distribution of particle, depth of shower maximum, kinetic energy distribution of particle at the ground, and energy deposited to the air. We then discuss implications of our results to the TA experiment.

\section{EAS simulation}

CORSIKA and COSMOS follow the development and evolution of EASs in the atmosphere; they describe the spatial, temporal, and energy distributions of secondary particles. To compare CORSIKA and COSMOS, we generated 50 EAS simulations with CORSIKA for each set of parameters (see below) and another 50 EAS simulations with COSMOS for each set of parameters. Primary energies of $E_{\rm 0}$ = 10$^{18.5}$eV, 10$^{18.75}$eV, 10$^{19}$eV, 10$^{19.25}$eV, 10$^{19.5}$eV, 10$^{19.75}$eV and 10$^{20}$eV were considered. And zenith angles of $\theta$ = 0\ensuremath{^\circ}, 18.2\ensuremath{^\circ}, 25.8\ensuremath{^\circ}, 31.75\ensuremath{^\circ}, 45\ensuremath{^\circ}, 70\ensuremath{^\circ} for proton and iron primaries were employed assuming the fat Earth. All together about 8,400 EASs were generated. Events with $\theta \le$ 45\ensuremath{^\circ} have been analyzed in the TA experiment \cite {Tinyakov-etal}, so we focus on the cases with $\theta \le$ 45\ensuremath{^\circ}. Simulations with $\theta$ = 70\ensuremath{^\circ} are used to estimate the energy deposited into the air in Section 3.4. Version 6.960 was used for CORSIKA simulations with $\theta$ = 0\ensuremath{^\circ}, 18.2\ensuremath{^\circ}, 25.8\ensuremath{^\circ}, 31.75\ensuremath{^\circ}, and 45\ensuremath{^\circ}, and version 6.980 was used for $\theta$ = 70\ensuremath{^\circ}. In COSMOS, for $\theta$ = 0\ensuremath{^\circ}, 18.2\ensuremath{^\circ}, 25.8\ensuremath{^\circ}, 31.75\ensuremath{^\circ}, and 45\ensuremath{^\circ}, 30 simulations were generated with version 7.54 and 20 simulations with version 7.581; the difference between versions 7.54 and 7.581 is small, so they were mixed. For $\theta$ = 70\ensuremath{^\circ}, version 7.581 was used.

\subsection{Interaction models of CORSIKA and COSMOS}

For high-energy ($E > 80$ GeV), the hadronic interaction generator QGSJETII-03 \cite{Ostapchenko} was used for both CORSIKA and COSMOS. QGSJETII-03 is one of most widely used interaction generators for UHECR EAS simulations.

For low-energy ($E < 80$ GeV), the hadronic interaction generator CORSIKA-FLUKA \cite{Battistoni-etal} was used in CORSIKA, while the Bertini and JQMD interaction models included in the PHITS code ($E < 2$ GeV) \cite{Niita-etal-phits} and JAM (v1.150) (2 GeV $< E <$ 80 GeV) \cite{Nara-etal} were used in COSMOS. We simply use the term ``PHITS'' for the two generators and the nelst routine managing  the elastic scattering.

In CORSIKA, the interactions of electro-magnetic (EM) particles (i.e., photons and electrons) were calculated with the EGS4 model \cite{Nelson-etal}. On the other hand, in COSMOS, the interactions were calculated with the Tsai's formula \cite{Tsai} and Nelson's formula \cite{Nelson-etal} which are based on the basic cross-sections of particles.

\subsection{Simulation set-up}

We employed the following simulation parameters, trying to compare CORSIKA and COSMOS simulations in parallel. First, in both CORSIKA and COSMOS, the Earth's magnetic field at the TA observation site (39.1\ensuremath{^\circ} N and 112.9\ensuremath{^\circ} W, just west of Delta, Utah) was used. The ground was fixed at the height of the TA site, 1430 m above the sea level, corresponding to the vertical atmospheric depth $x_v = 875$ g/cm$^{2}$. Second, in both CORSIKA and COSMOS, the same threshold energies, $E_{\rm threshold}$, were applied to secondary particles. Particles having energy below $E_{\rm threshold}$ were not tracked in simulations. $E_{\rm threshold}$ = 500 keV was used for EM particles, while $E_{\rm threshold}$ = 50 MeV for muons and hadrons. Most particles reach the ground with energy larger than the $E_{\rm threshold}$ (see Figures 4 and 5 for the case of iron primary with $E_{\rm 0}$ = 10$^{19.5}$eV and $\theta$ = 0\ensuremath{^\circ}). Third, the Landau-Pomeranchuk-Migdal (LPM) effect \cite{Landau-etal, Migdal1956, Alvarez-Muniz1998} is included in both CORSIKA and COSMOS. The LPM effect causes a reduction of bremsstrahlung and pair production cross sections at high energies.

In principle, all secondary particles can be tracked along their trajectories and their physical properties can be stored until they reach the ground. Then, the number of particles can become too large to be comfortably accommodated with available computational resources. To alleviate this problem, most EAS simulations introduce the Hillas thinning algorithm \cite{Hillas1997}. The algorithm picks up only a small fraction for secondary particles with energy smaller than the product of the primary energy ($E_{\rm 0}$) and a thinning level ($L_{\rm th})$, i.e., for particles with $E \le E_{\rm 0} \times L_{\rm th}$. At each vertex of interaction, one secondary particle is selected in a way that more energetic particles are picked up with higher probabilities, and further tracked. A weight, which is defined as the ratio of the energy of the selected particle to that of all secondary particles at the vertex, is assigned to the selected particle to represent untracked particles. Eventually, the total number of secondary particles are recovered by counting tracked particles multiplied by their weights \cite{Kobal}. 

In CORSIKA, it is recommended to take a value between $10^{-3}$ and $10^{-7}$ for $L_{\rm th}$. We chose $L_{\rm th} = 10^{-7}$. On the other hand, in COSMOS, it is recommended to use $L_{\rm th}=A \times 10^{-7}$, where $A$ is the mass number. Hence, for proton primary, $L_{\rm th} = 1 \times 10^{-7}$ was used for both CORSIKA and COSMOS; for iron primary, $L_{\rm th} = 1 \times 10^{-7}$ and $L_{\rm th} = 5.6 \times 10^{-6}$ were used for CORSIKA and COSMOS, respectively. In addition, COSMOS applies a smaller thinning, $L'_{\rm th}= 10^{-2} \times L_{\rm th}$, where a higher accuracy is required. At the upper atmosphere of $x_v <400$ g/cm$^{2}$ or near the shower core of $r<20$ m, $L_{\rm th}$ is used, while $L'_{\rm th}$ is applied to the region of $x_v = 400 - 875$ g/cm$^{2}$ and $r \ge 20$ m.

CORSIKA and COSMOS both have an upper limit on the weight, the so-called maximum weight value, $W_{\rm max}$. As an EAS develops, through interactions of particles, weights of tracked particles are continuously accumulated. When accumulated weights reach $W_{\rm max}$, the thinning algorithm is no longer applied, and particles are tracked without further thinning. We used $W_{\rm max}$ which is differently for CORSIKA and COSMOS: $W_{\rm max}=L_{\rm th} \times (E_{\rm 0} [{\rm eV}] /10^{9})$ for CORSIKA \cite{Knapp} and $W_{\rm max}=E_{\rm 0} [{\rm eV}]/10^{15}$ for COSMOS \cite{Cohen-Kasahara}. 

With our choices of thinning and weighting, overall more particles are tracked in COSMOS than in CORSIKA. The computation time is roughly proportional to the number of tracked particles, so for the same primary COSMOS simulations presented here took longer than CORSIKA simulations. 

We also note that the data dumping is different between CORSIKA and COSMOS. In CORSIKA, the grid points of the vertical atmospheric depth have a spacing of $\Delta x_v =$ 1 g/cm$^{2}$. On the other hand, in COSMOS, the grid points are defined at $x_v = 0$, 100, 200 g/cm$^{2}$, and after $200$ g/cm$^{2}$ they have a spacing of $\Delta x_v =$ 25 g/cm$^{2}$. So the data from CORSIKA simulations are dumped in every $\Delta x_v =$ 1 g/cm$^{2}$, while the data from COSMOS are dumped in every $\Delta x_v =$ 100 g/cm$^{2}$ for $x_v \leq$ 200 g/cm$^{2}$ and in every $\Delta x_v =$ 25 g/cm$^{2}$ for $x_v >$ 200 g/cm$^{2}$.

\section{Comparison of CORSIKA and COSMOS simulation results}

\subsection{Longitudinal distribution of particles}

When UHECRs strike the atmosphere, most of the particles initially generated are neutral and charged-pions. Neutral-pions quickly decay into two photons. Charged-pions (positively or negatively charged) survive longer, and either collide with other particles or decay to muons and muon neutrinos. Those particles produce the so-called EM and hadronic showers. In EM showers, photons create electrons and positrons by pair-production, and in turn electrons and positrons create photons via bremsstrahlung, and so on. EM showers continue until the average energy per particle drops to $\sim 80$ MeV. Below this energy, the dominant energy loss mechanism is ionization rather than bremsstrahlung. Then, EM particles are not efficiently produced anymore, and EASs reach the maximum (see the next subsection). In hadronic showers, muons and hadrons are produced through hadronic interactions and decays. Here, hadrons include nucleons (neutrons and protons), pions, and kaons.

The number of secondary particles created by EM and hadronic showers initially increases and then decreases, as an EAS develops through the atmosphere. The distribution of particles along the atmospheric depth is called the longitudinal distribution \cite{Bock-etal,Pryke}. Here, we first compare the longitudinal distributions from CORSIKA and COSMOS simulations, and analyze the differences in photon, electron, muon, and hadron distributions.

Figures 1 and 2 show the typical longitudinal distributions as a function of slant atmospheric depth, $x_s = x_v / \cos\theta$. Lines represent the numbers of particles averaged for 50 EAS simulations, $\left< N \right>$, and error bars mark the standard deviations, $\sigma$, defined as 
\begin{equation}
\sigma = \sqrt{\frac {1}{n_{\rm sim}} \sum\limits_{i=1}^{n_{\rm sim}} (N_{i}-\left< N \right>)^2}.
\end{equation}
Here, $n_{\rm sim}=50$ is the number of EAS simulations for each set of parameters and $N_i$ is the number of particles  at $x_s$ in each simulation. The EASs shown are for proton and iron primaries, respectively, with $E_{\rm 0}$=10$^{19.5}$ eV and $\theta$ = 0\ensuremath{^\circ} and 45\ensuremath{^\circ}. Numbers for photons, electrons, muons, and hadrons are shown. Table 1 shows the numbers of particles at peaks, which are again averages of 50 EAS simulations. If the peaks are located beyond the maximum depth, the values at the maximum depth are shown. Note that difference species have peaks at different $x_s$'s.

In the cases shown, COSMOS predicts slightly more particles in the early stage of EASs (except for photons in the upper-left panel of Figure 1), while CORSIKA predicts slightly more particles in the late stage. But the difference is within the fluctuation (that is, less than $\sigma$ in Equation (1)). Quantitatively, the difference between CORSIKA and COSMOS results in the peak numbers of particles is at most $7 - 8 \%$, as shown in Table 1.

\subsection{Depth of shower maximum}

The depth of shower maximum, denoted by $X_{\rm max}$, is defined as the slant atmospheric depth at which the number of secondary electrons reaches the maximum in EASs. $X_{\rm max}$ is a function of the primary energy, but it has different values and dispersions for different primary particles. For a given primary energy, proton primary has larger values and dispersions of $X_{\rm max}$ than iron primary. The average and standard deviation of $X_{\rm max}$, $\left<X_{\rm max}\right>$ and $\sigma_{X_{\rm max}}$, are known as the key quantities that discriminate the composition of primary particles in UHECR experiments.

In calculating $X_{\rm max}$, the longitudinal distribution of electrons along the shower axis was fitted to the Gaisser-Hillas function (GHF) \cite{Gaisser},
\begin{equation}
N_{\rm electron}(x_s) = N_{\rm electron, max}\left(\frac {x_s-x_{s0}} {X_{\rm max}-x_{s0}} \right)^{\frac {X_{\rm max}-x_{s0}} {\lambda}} \exp \left({\frac {X_{\rm max}-x_s} {\lambda}}\right),
\end{equation}
where $N_{\rm electron, max}$ is the maximum number of electrons at $X_{\rm max}$. We sought $X_{\rm max}$ by treating $x_{s0}$ and $\lambda$ as well as $X_{\rm max}$ and $N_{\rm electron, max}$ as fitting parameters. We note that originally $x_{s0}$ was meant to be the depth at which the first interaction occurs and $\lambda$ to be the proton interaction mean free path. But in practice, they were regarded as fitting parameters. It was shown that the resulting $X_{\rm max}$ is not sensitive to whether $\lambda$ is used as a free parameter or set to a fixed value \cite{Song, Abbasi-etal-5}.

Figure 3 and Table 2 show $\left<X_{\rm max}\right>$ and $\sigma_{X_{\rm max}}$ in our simulations for proton and iron primaries with different $E_0$'s. $\left<X_{\rm max}\right>$ and $\sigma_{X_{\rm max}}$ in Table 2 were calculated for 250 simulations including those of five different zenith angles ($\theta \leq 45\ensuremath{^\circ}$). Solid and dashed lines in Figure 3 are the least chi-square fits of $\left<X_{\rm max}\right>$ and $\sigma_{X_{\rm max}}$ in Table 2. The result of Wahlberg et al. with CORSIKA \cite{Wahlberg-etal} is included with dot-dashed lines for comparison.

We note that in some simulations the shower maximum occurred beyond the maximum depth. In such cases, $X_{\rm max}$'s from fits to the GHF may have larger errors. And for CORSIKA results, the data dumped in every $\Delta x_v =$ 1 g/cm$^{2}$ were used, while for COSMOS results, the data in every $\Delta x_v =$ 25 g/cm$^{2}$ were used. So a larger systematic error may exist in COSMOS results.

The results for $\left<X_{\rm max}\right>$ and $\sigma_{X_{\rm max}}$ in Figure 3 and Table 2 are summarized as follows. First, the difference between CORSIKA and COSMOS results in $\left<X_{\rm max}\right>$ is at most $\sim 16$ g/cm$^{2}$ for both proton and iron primaries. It is smaller than the fluctuation, $\sigma_{X_{\rm max}}$. Second, the difference between $\left<X_{\rm max}\right>$'s for proton and iron primaries is typically $\sim 70 - 80$ g/cm$^{2}$, which is beyond the fluctuations both in CORSIKA and COSMOS simulations as well as the difference between CORSIKA and COSMOS results. Third, $\sigma_{X_{\rm max}}$ is $\sim 40 - 60 $ g/cm$^{2}$ in for proton primary, while it is $\sim 20 - 25 $ g/cm$^{2}$ for iron primary. $\sigma_{X_{\rm max}}$ is somewhat larger in CORSIKA than in COSMOS, as is clear in Figure 3; the difference is larger for proton primary. Fourth, our CORSIKA results agree with those of Wahlberg et al. Yet ours are smaller by up to $\sim 10$ g/cm$^{2}$. A number of possible causes can be conjectured. Our simulations performed with versions, models, and parameters different from those of Wahlberg et al. In our work $\left<X_{\rm max}\right>$ is defined as the depth of the peak in the number of electrons above 500 keV, while in Wahlberg et al. it was defined as the depth of the peak in overall energy deposit. Also the error in the fitting could be in the level of $\sim 10$ g/cm$^{2}$. Although not shown here, we found that $\left<X_{\rm max}\right>$ for different zenith angles varies by up to $\sim 10$ g/cm$^{2}$.

\subsection{Kinetic energy distribution of particles at the ground}

In EASs, a fraction of secondary particles reach the ground. Those particles deposit a part of their energy to ground detectors, such as scintillation detectors or water Cherenkov tanks. In experiments, by measuring the amount and spatial distribution of the deposited energy, the primary energy and arrival direction of UHECRs are estimated \cite{Hillas-etal}. Here, we present the kinetic energy (i.e., the total energy subtracted by the rest-mass energy) distributions of secondary particles over the entire ground; the amount of energy deposited to detectors is determined by the kinetic energy.

Figure 4 shows the typical kinetic energy distributions of photons, electrons, muons, and hadrons, including particles in the shower core; here the EAS is for iron primary with $E_{\rm 0} = 10^{19.5}$ eV and $\theta = 0\ensuremath{^\circ}$. Lines are the averages of 50 EAS simulations, and error bars mark the standard deviations, $\sigma$, defined similarly as in Equation (1). Tables 3, 4, 5, and 6 show the total kinetic energies ($E$) and numbers ($N$) of particles reaching the ground for each particle species. Again, they are the averages of 50 EAS simulations. To further analyze the kinetic energy distributions of different components, hadrons were separated into nucleons, pions, and kaons, and Figure 5 shows their distributions.

We first point that although $N_{\rm photon} \gg N_{\rm electron} \gg N_{\rm muon} \gg N_{\rm hadron}$ for all the cases we simulated as shown in Tables 5 and 6, the energy partitioning depends on EAS parameters and varies significantly as shown in Tables 3 and 4. For instance, in the EAS of iron primary with $E_0 = 10^{19.5}$ eV and $\theta = 0\ensuremath{^\circ}$ which is shown in Figures 4 and 5, the partitioning of the kinetic energies of particles reaching the ground is $E_{\rm EM} : E_{\rm muon} : E_{\rm hadron} \sim 1 : 0.18 : 0.11$. On the other hand, in the EAS of proton primary with $E_0 = 10^{18.5}$ eV and $\theta = 45\ensuremath{^\circ}$, $E_{\rm EM} : E_{\rm muon} : E_{\rm hadron} \sim 1 : 1.1 : 0.11$.

We found that the difference between CORSIKA and COSMOS results in Figures 4 and 5 is up to 30 $\%$, but yet the difference is within the fluctuation at most energy bins. Tables 3, 4, 5, and 6 indicate differences of up to 30 $\%$ in the integrated kinetic energies and numbers. There are following general tends: 1) For most cases, CORSIKA predicts larger energies for photons and electrons, while COSMOS predicts larger energies for muons. 2) The difference is larger for proton primary than for iron primary. 3) The difference is larger for larger $E_0$ and for larger $\theta$. We note that larger numbers of particles do not necessarily mean larger energies; this point is particularly clear for muons.

\subsection{Energy deposited to the air}

Interactions between air molecules and secondary particles yield UV fluorescence light, which is observed with fluorescence telescopes in UHECR experiments \cite{Song-etal,Fabjan-Gianotti}. The energy estimated through observation of UV fluorescence light is called the calorimetric energy, and it is used to infer the primary energy of UHECRs \cite{Barbosa-etal}. The energy released as the fluorescence light is determined by the energy deposited to the air, $E_{\rm air}$. So in order for the primary energy to be accurately estimated in UHECR experiments, $E_{\rm air}$ needs to be precisely known \cite{M.Risse-D.Heck}. 

We compare the energy deposited to the air due to EM particles, muons, and hadrons in CORSIKA and COSMOS simulations. Both CORSIKA and COSMOS follow $E_{\rm air}$ by the particles with $E > E_{\rm threshold}$ along the atmospheric depth in simulations. But the codes does not track particles and their contribution any more, if their energy drops below the threshold energy. So we compare $E_{\rm air}$ by particles with $E > E_{\rm threshold}$. Figure 6 shows $E_{\rm air}$ as a function of the slant atmospheric depth, $x_s$, for proton primary with $E_{\rm 0} = 10^{19.5}$ eV and $\theta = 0\ensuremath{^\circ}$, $31.75\ensuremath{^\circ}$, $45\ensuremath{^\circ}$ and $70\ensuremath{^\circ}$. Lines are the averages of 50 EAS simulations. Table 7 shows the average of the fraction of the energy, $\left<E_{\rm air}\right>/E_{0}$, and the relative standard deviation, $\sigma_{E_{\rm air}}$/$\left<E_{\rm air}\right>$, for proton and iron primaries with different primary energies and $\theta = 70\ensuremath{^\circ}$ at the ground; for $\theta = 70\ensuremath{^\circ}$ the slant atmospheric depth at the ground is large enough that $E_{\rm air}$ has reached the maximum (see Figure 6). The values in Table 7 were calculated with 50 EAS simulations for each set of parameters.

There is a clear trend in Figure 6 that COSMOS predicts larger $E_{\rm air}(x_s)$ than CORSIKA. The energy deposited to the air by the particles with $E > E_{\rm threshold}$ in Table 7 is $\left<E_{\rm air}\right>$/$E_{\rm 0} = 0.66 - 0.71$ in CORSIKA simulations, while $\left<E_{\rm air}\right>$/$E_{\rm 0} = 0.77 - 0.82$ in COSMOS simulations. The difference is $\sim 15$ $\%$, which is larger than the fluctuation. The relative standard deviation, $\sigma_{E_{\rm air}}$/$\left<E_{\rm air}\right>$, is small and typically $\sim 1$ $\%$ both in CORSIKA and COSMOS simulations.

For the total energy deposited to the air, the contribution due to the particles with $E < E_{\rm threshold}$, as well as that by the particles with $E > E_{\rm threshold}$, should be counted. Yet, the difference of $\sim 15$ $\%$ is substantial. It means that the UV fluorescence light assessed with CORSIKA and COSMOS simulations could differ by a similar amount, so does the primary energy of UHECR events estimated with CORSIKA and COSMOS simulations.

\section{Summary}

EAS simulations form an essential part of experiments to detect UHECRs; they are used to estimate the energy, arrival direction, and composition of primary particles. The TA experiment employs two codes, CORSIKA and COSMOS, for the simulations. In this paper, we compared CORSIKA and COSMOS simulations by quantifying the differences in the longitudinal distribution of particles, depth of shower maximum, kinetic energy distribution of particle at the ground, and energy deposited to the air.

Most of all, we should point that the simulation results of CORSIKA and COSMOS agree well with each other, despite of all the complexities and differences in the models and elements involved in the codes. Such agreement should be quite an achievement. Nevertheless, there are non-negligible differences in the quantities we presented. Those may be regarded as systematic uncertainties in the EAS simulation part of UHECR experiments.

1) The difference between CORSIKA and COSMOS results in the longitudinal distribution of particles is less than $\sim 10$ $\%$ for most cases, which is within the fluctuation. The difference in the peak numbers of particles is at most $7 - 8$ $\%$ (Table 1). COSMOS tends to predict slightly more particles in the early stage of EASs, while CORSIKA tends to predict slightly more particles in the late stage.

2) CORSIKA and COSMOS predict the depths of shower maximum, $\left<X_{\rm max}\right>$, which are consistent with each other. The difference between $\left<X_{\rm max}\right>$'s from CORSIKA and COSMOS is up to $\sim 16$ g/cm$^{2}$, which is noticeable but smaller than the fluctuation, $\sigma_{X_{\rm max}}$. The difference between $\left<X_{\rm max}\right>$'s for proton and iron primaries, which is typically $\sim 70 - 80$ g/cm$^{2}$, is beyond the fluctuations both in CORSIKA and COSMOS simulations as well as the difference between the CORSIKA and COSMOS results. $\sigma_{X_{\rm max}}$ is $\sim 40 - 60 $ g/cm$^{2}$ in for proton primary, while it is $\sim 20 - 25 $ g/cm$^{2}$ for iron primary. $\sigma_{X_{\rm max}}$ is somewhat larger in CORSIKA than in COSMOS; the difference is larger for iron primary.

3) There are differences of up to 30 $\%$ between CORSIKA and COSMOS results in the numbers and energies of the particles reaching the ground; the difference is larger for proton primary with larger $E_0$'s and larger $\theta$'s. It implies that the amount of the energy deposited to ground detectors could be different up to 30 $\%$ or so in CORSIKA and COSMOS simulations. The exact response to the particles passing through ground detectors, however, depends on the details of detectors, and need simulations, for instance, with the GEANT code. We leave this issue for a follow-up study.

4) The energy deposited to the air, $E_{\rm air}$, by the particles with $E > E_{\rm threshold}$ is larger by $\sim 15$ $\%$ in COSMOS simulations than in CORSIKA simulations. This implies that the UV fluorescence light assessed with CORSIKA and COSMOS simulations could differ by a similar amount.

\section*{Acknowledgments}

We thank the referee for critical reading of the manuscript and constructive comments. We thanks the CORSIKA user support team for help in using CORSIKA. KK thanks K. Niita for courtesy and help in implementing the PHITS code. The work of SR, JK, DR, and HK was supported by the National Research Foundation of Korea through grant 2007-0093860. The work of KK, EK, and AT was supported by Grant-in-Aid for Scientific Research on Priority Areas (Highest Energy Cosmic Rays: 15077205).

\clearpage

\begin{figure}
\hskip -0.6cm
\centerline{\includegraphics[width=150mm]{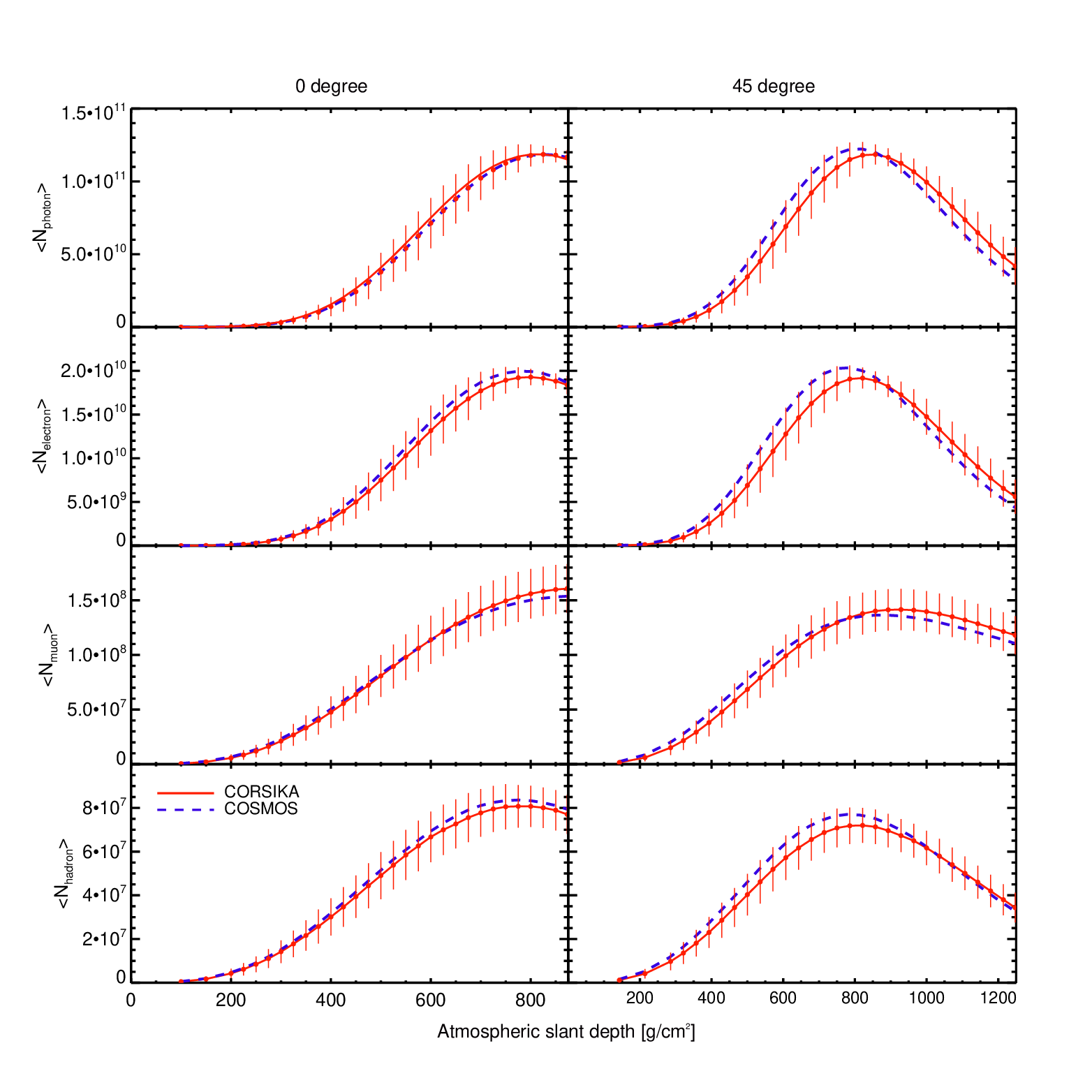}}
\caption{Longitudinal distribution of photons, electrons, muons, and hadrons for EASs of proton primary with $E_{\rm 0}$ = 10$^{19.5}$ eV and $\theta=$ 0\ensuremath{^\circ} (left panels) and 45\ensuremath{^\circ} (right panels). Lines represent the averages of 50 simulations, and error bars mark the standard deviations. For clarity, only the error bars of CORSIKA results are shown.}
\end{figure}

\clearpage

\begin{figure}
\hskip -0.6cm
\centerline{\includegraphics[width=150mm]{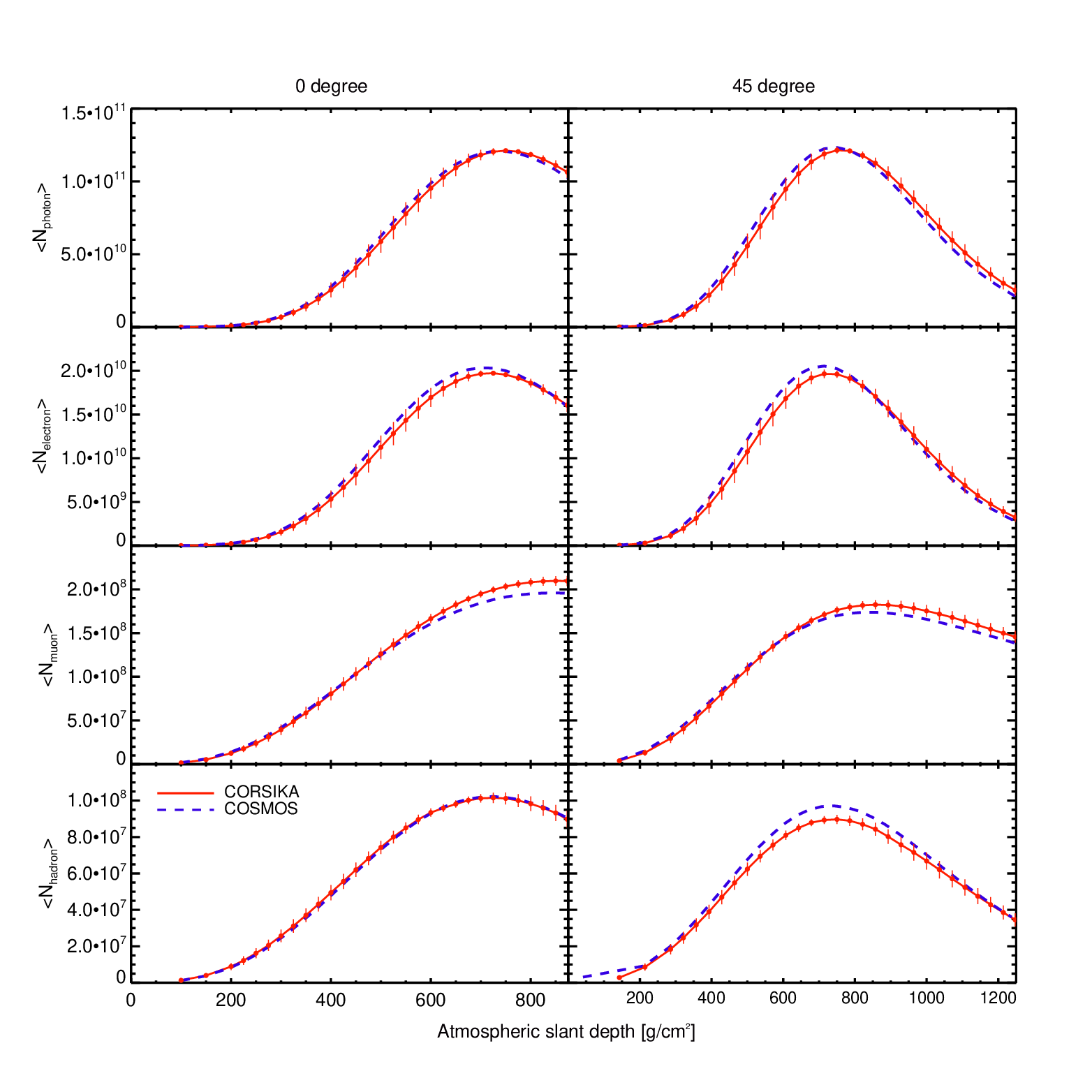}}
\caption{Longitudinal distribution of photons, electrons, muons, and hadrons for EASs of iron primary with $E_{\rm 0}$ = 10$^{19.5}$ eV and $\theta=$ 0\ensuremath{^\circ} (left panels) and 45\ensuremath{^\circ} (right panels). Lines represent the averages of 50 simulations, and error bars mark the standard deviations. For clarity, only the error bars of CORSIKA results are shown.}
\end{figure}

\clearpage
\begin{figure}
\hskip -0.6cm
\centerline{\includegraphics[width=155mm]{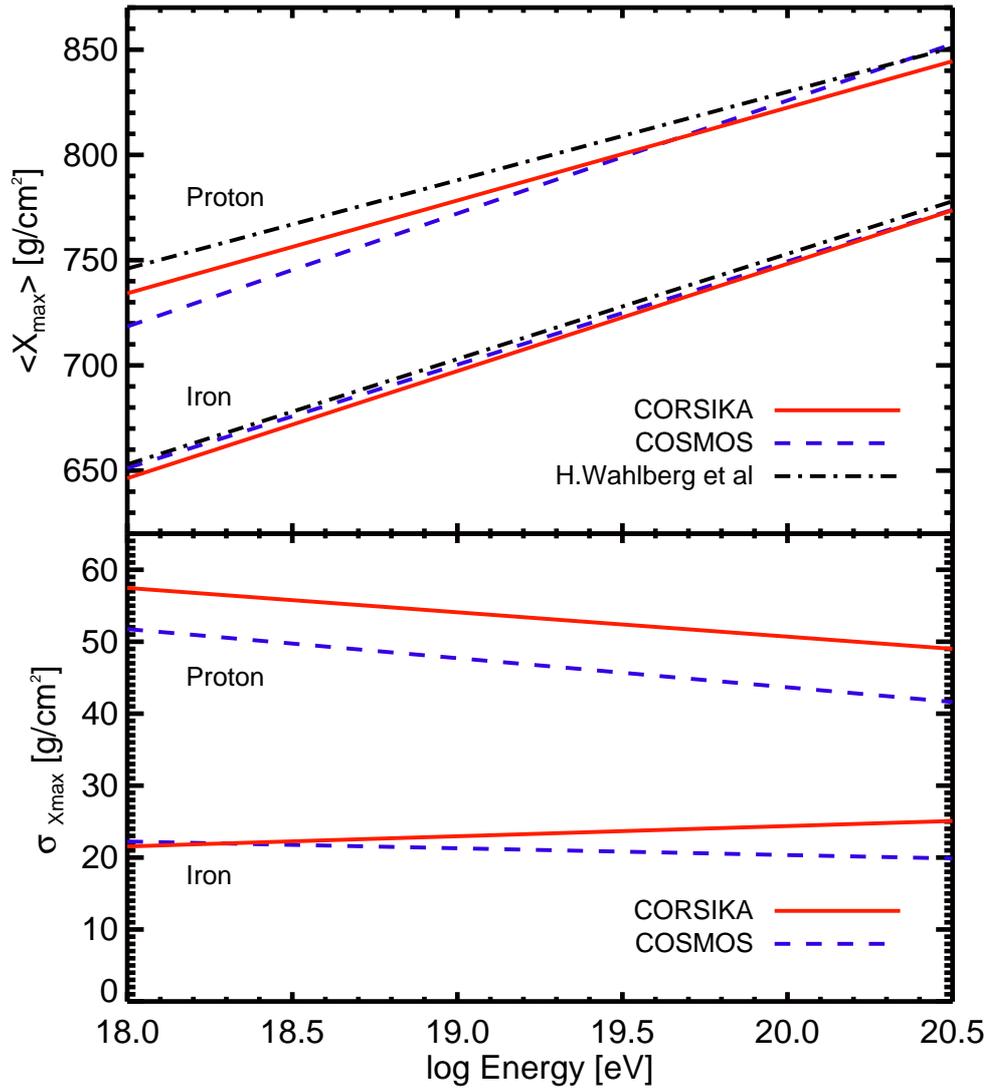}}
\caption{Average of shower maximum, $\left<X_{\rm max}\right>$ (upper panel), and standard deviation, $\sigma_{X_{\rm max}}$ (lower panel,) as a function of primary energy. Lines are the least chi-square fits of the values in Table 2, which were calculated for 250 simulations for all zenith angles. The result reported in \cite{Wahlberg-etal} is included for comparison.}
\end{figure}

\clearpage

\begin{figure}
\hskip -0.3cm
\centerline{\includegraphics[width=190mm]{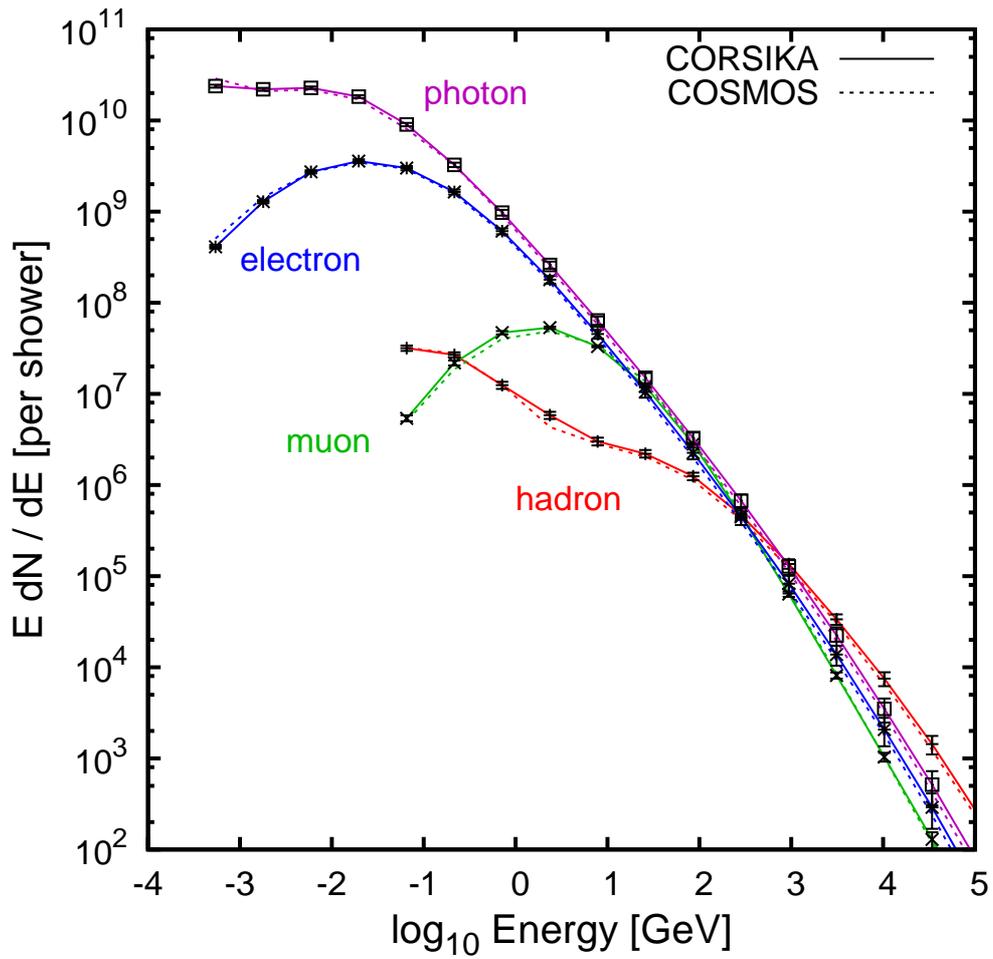}}
\caption{Kinetic energy distribution of photons, electrons, muons, and hadrons at the ground for EASs of iron primary with $E_{\rm 0}$ = 10$^{19.5}$eV and $\theta$= 0\ensuremath{^\circ}. Solid lines are CORSIKA results and dashied lines are COSMOS. Violet, blue, green and red colors indicate photons, electrons, muons and hadrons, respectively. Lines are the averages of 50 simulations, and error bars mark the standard deviations. For clarity, only the error bars of CORSIKA results are shown.}
\end{figure}

\clearpage

\begin{figure}
\hskip -0.3cm
\centerline{\includegraphics[width=190mm]{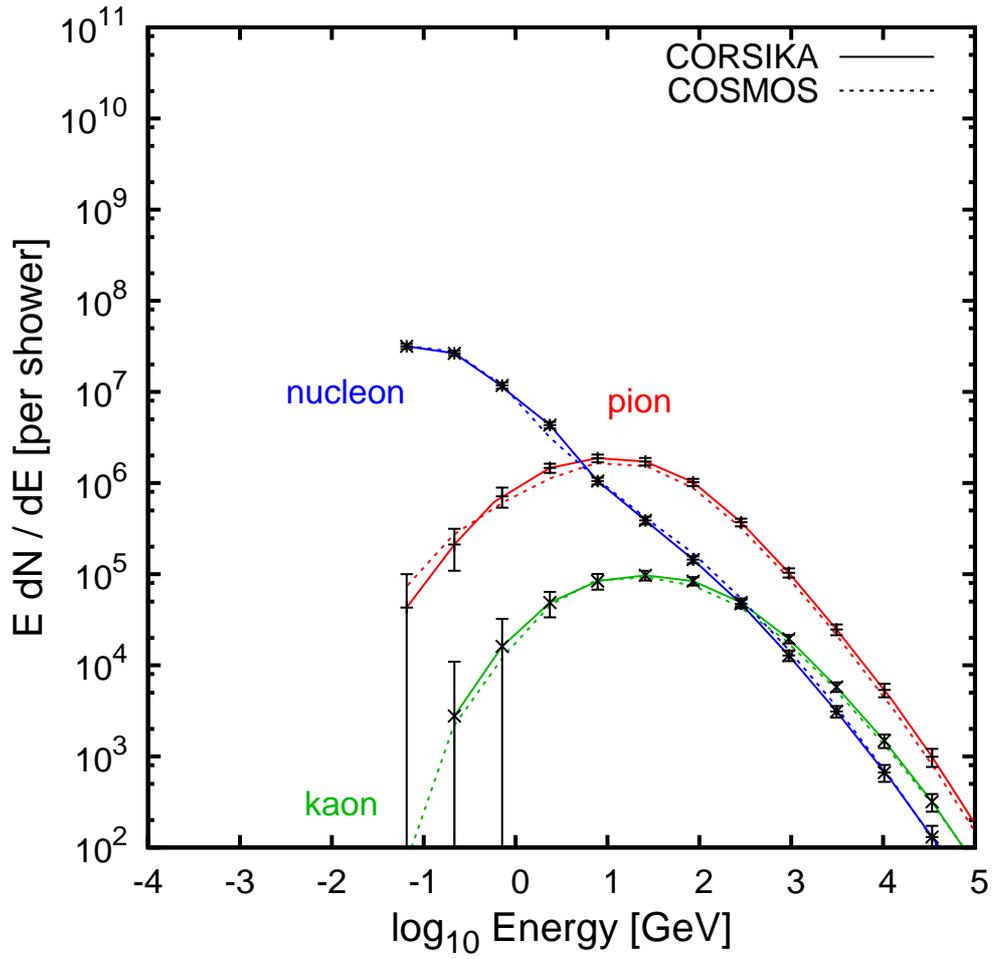}}
\caption{Kinetic energy distribution of nucleons, pions, and kaons at the ground for EASs of iron primary with $E_{\rm 0}$ = 10$^{19.5}$eV and $\theta$= 0\ensuremath{^\circ}. Solid lines are CORSIKA results and dashied lines are COSMOS. Blue, red and green colors indicate nucleons, pions and kaons, respectively. Lines are the averages of 50 simulations, and error bars mark the standard deviations. For clarity, only the error bars of CORSIKA results are shown.}
\end{figure}

\clearpage

\begin{figure}
\hskip -0.5cm
\centerline{\includegraphics[width=135mm]{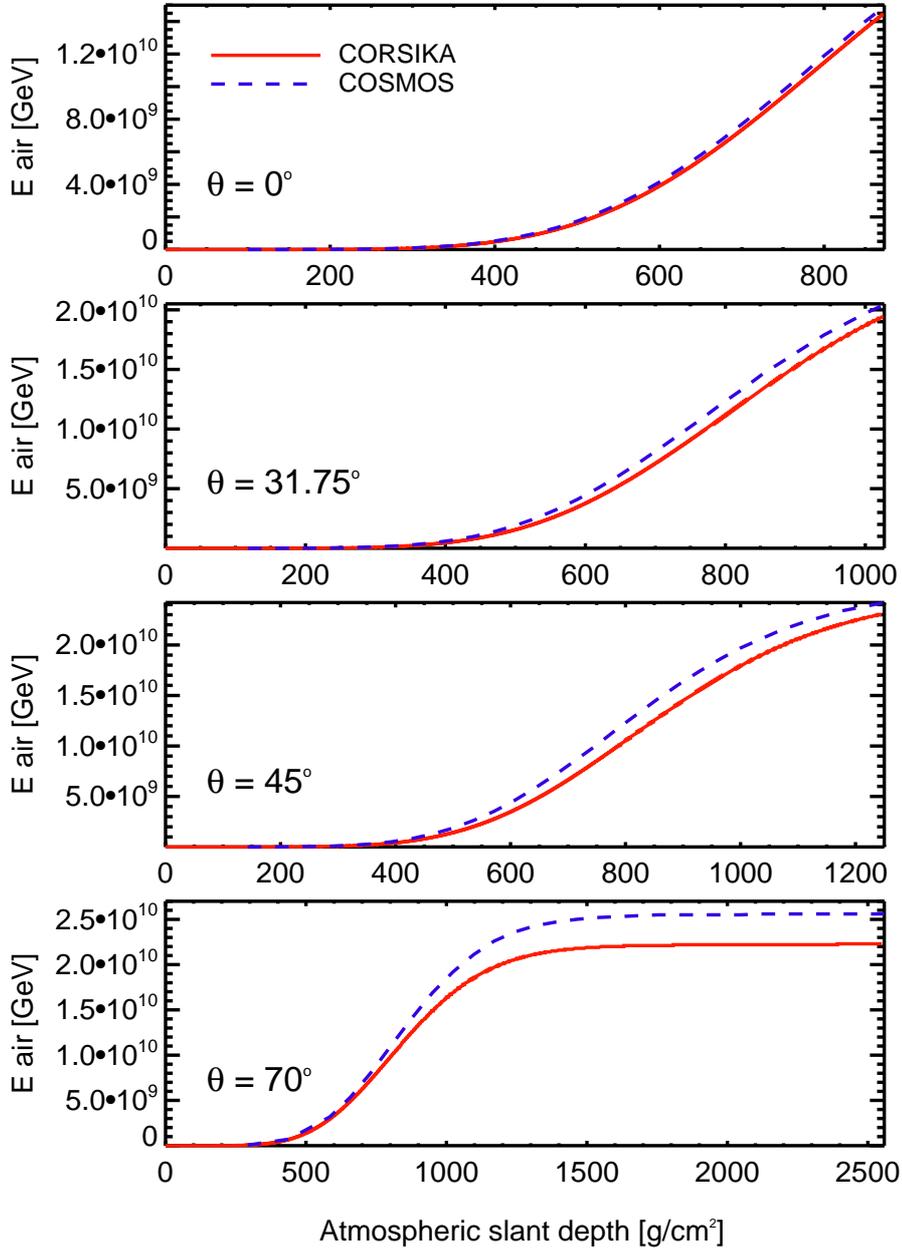}}
\caption{Energy deposited to the air by the particles with energy above the threshold energy as a function of slant atmospheric depth, $x_s$, for EASs of proton primary with $E_{\rm 0}$ = 10$^{19.5}$eV. Shown are for $\theta$ = 0\ensuremath{^\circ}, $\theta$ = 31.75\ensuremath{^\circ}, $\theta$ = 45\ensuremath{^\circ}, and $\theta$ = 70\ensuremath{^\circ} from top to bottom. Lines are the averages of 50 simulations.}
\end{figure}

\clearpage

\begin{table}
\begin{center}
\begin{tabular}[t]{c||c|cccccc}\hline
\multicolumn{6}{c}{Number of particles at peak}\\\hline
primary / $\theta$ & simulation & photons & electrons & muons &
hadrons  \\\hline\hline
proton& CORSIKA &  1.19e+11 & 1.93e+10 & 1.61e+08 & 8.07e+07 \\
0\ensuremath{^\circ}& COSMOS & 1.19e+11 & 2.00e+10 & 1.54e+08 &
8.36e+07 \\\cline{1-6}
proton& CORSIKA &  1.18e+11 & 1.92e+10 & 1.42e+08& 7.20e+07 \\
45\ensuremath{^\circ}& COSMOS &  1.22e+11 & 2.04e+10 & 1.36e+08&
7.30e+07 \\\hline\hline
Iron& CORSIKA &   1.21e+11 &1.97e+10 & 2.10e+08 & 1.02e+08 \\
0\ensuremath{^\circ}& COSMOS &  1.21e+11 & 2.04e+10 & 1.96e+08 &
1.02e+08 \\\cline{1-6}
Iron& CORSIKA &  1.21e+11 & 1.97e+10 & 1.83e+08 & 8.97e+07 \\
45\ensuremath{^\circ}& COSMOS & 1.23e+11 & 2.05e+10 & 1.74e+08 &
9.71e+07 \\\hline
\end{tabular}
\caption{Number of particles at peak for proton and iron primaries with $E_{\rm 0}$ = 10$^{19.5}$ eV and $\theta$ = 0\ensuremath{^\circ} and 45\ensuremath{^\circ}. Averages of 50
simulations are listed.}
\end{center}
\end{table}

\begin{table}
\begin{center}
\begin{tabular}[t]{c||c|ccccccccc}\hline
\multicolumn{9}{c}{Depth of shower maximum, $X_{\rm max}$ (units: g/cm$^2$)}\\\hline
primary& $\log_{10}E_0$ (eV) &18.5&18.75&19&19.25&19.5&19.75&20\\\hline\hline
&CORSIKA $\left<X_{\rm max}\right>$& 754.1 & 768.7 & 779.5 & 789.3 & 802.1 & 810.3 & 821.8\\
proton& $\sigma_{X_{\rm max}}$ & 52.6& 59.4 & 55.1 & 49.0 & 55.8 & 50.0 & 50.7\\\cline{2-9}
&COSMOS $\left<X_{\rm max}\right>$& 746.2 & 760.0 & 774.9 & 781.2 & 781.3 & 813.8 & 836.8\\
& $\sigma_{X_{\rm max}}$ & 46.8 & 45.2 & 53.1 & 50.2 & 48.5 &42.2 & 40.9 \\\hline\hline
&CORSIKA $\left<X_{\rm max}\right>$& 672.5 & 682.2 & 698.0 &711.8 &722.3 &735.8 &747.6\\
iron& $\sigma_{X_{\rm max}}$ & 23.1 & 20.9 & 23.6 & 23.4 & 23.5 & 25.2 & 23.6\\\cline{2-9}
&COSMOS $\left<X_{\rm max}\right>$& 671.9& 698.6 & 704.9 & 702.8 & 713.0 & 742.7 & 754.4 \\
& $\sigma_{X_{\rm max}}$ & 19.5 & 24.6 & 20.5 & 23.2 & 19.0 & 18.7 & 21.8 \\\hline
\end{tabular}
\caption{Average and standard deviation of $X_{\rm max}$, which were calculated for 250 simulations for all zenith angles.}
\end{center}
\end{table}

\clearpage

\begin{table}
\begin{center}
\begin{tabular}[t]{c|c|c| c cccccc}\hline
\multicolumn{8}{c}{Kinetic energies of particles at the ground for proton primary (units: GeV)}\\\hline
$\theta$&$\log_{10}E_0$ (eV) & simulation & photons & electrons & muons & hadrons & all \\\hline\hline
&18.5 & CORSIKA & 6.56e+08 & 3.71e+08 & 1.26e+08 & 8.75e+07 & 1.24e+09\\
& &COSMOS & 6.17e+08 & 3.46e+08 & 1.45e+08 & 9.07e+07 & 1.20e+09\\\cline{2-8}
0\ensuremath{^\circ} &19 & CORSIKA & 2.29e+09 & 1.32e+09 & 3.51e+08 & 2.70e+08 & 4.24e+09\\
& &COSMOS & 2.18e+09 & 1.26e+09 & 3.96e+08 & 2.73e+08 & 4.11e+09\\\cline{2-8}
&19.5 & CORSIKA & 7.91e+09 & 4.68e+09 & 1.01e+09 & 8.44e+08 & 1.44e+10\\
& &COSMOS & 7.40e+09 & 4.35e+09 & 1.11e+09 & 7.83e+08 & 1.36e+10\\\cline{1-8}
\hline\hline
&18.5 & CORSIKA & 9.71e+07 & 4.20e+07 & 1.33e+08 & 1.32e+07 & 2.85e+08\\
& &COSMOS & 8.74e+07 & 3.89e+07 & 1.59e+08& 1.56e+07 & 3.01e+08\\\cline{2-8}
45\ensuremath{^\circ} &19&CORSIKA& 4.04e+08 & 1.88e+08 & 3.89e+08 & 4.74e+07 & 1.03e+09\\
& &COSMOS & 3.48e+08 & 1.58e+08 & 4.32e+08 & 4.46e+07 & 9.83e+08\\\cline{2-8}
&19.5 & CORSIKA  & 1.54e+09 & 7.31e+08 & 1.13e+09 & 1.47e+08 & 3.55e+09\\
& &COSMOS & 1.09e+09 & 5.02e+08 & 1.24e+09 & 1.21e+08 & 2.95e+09\\\cline{1-8}
\end{tabular}
\caption{Total kinetic energy of particles reaching the ground, for proton primary. Averages of 50 simulations are listed.}
\end{center}
\end{table}

\begin{table}
\begin{center}
\begin{tabular}[t]{c|c||c| cccccc}\hline
\multicolumn{8}{c}{Kinetic energies of particles at the ground for iron primary (units: GeV)}\\\hline
$\theta$&$\log_{10}E_0$ (eV) & simulation & photons & electrons & muons & hadrons & all \\\hline\hline
&18.5 & CORSIKA & 4.52e+08 & 2.38e+08 & 1.89e+08 & 9.54e+07 & 9.75e+08\\
&& COSMOS & 4.26e+08 & 2.23e+08 &1.99e+08 & 9.24e+07 & 9.40e+08\\\cline{2-8}
0\ensuremath{^\circ}&19 & CORSIKA & 1.59e+09 &8.54e+08 &5.25e+08 &2.94e+08 &3.27e+09\\
&& COSMOS & 1.63e+09 &8.81e+08 &5.43e+08 &2.94e+08 &3.35e+09\\\cline{2-8}
&19.5 & CORSIKA & 5.70e+09 & 3.12e+09 & 1.45e+09 & 9.04e+08 & 1.12e+10\\
&& COSMOS & 5.30e+09 & 2.89e+09 & 1.54e+09 & 9.14e+08 & 1.06e+10\\\cline{1-8}
&18.5 & CORSIKA & 6.00e+07 & 2.61e+07 & 2.01e+08 & 1.40e+07 & 3.01e+08\\
&& COSMOS & 5.17e+07 & 2.25e+07 & 2.14e+08 & 1.36e+07 & 3.02e+08\\\cline{2-8}
45\ensuremath{^\circ}& 19 & CORSIKA & 2.26e+08 & 9.96e+07 & 5.62e+08 & 4.59e+07 & 9.34e+08\\
&& COSMOS & 2.03e+08 & 8.87e+07 & 5.89e+08 & 4.34e+07 & 9.24e+08\\\cline{2-8}
&19.5 & CORSIKA & 8.35e+08 & 3.70e+08 & 1.57e+09 & 1.44e+08 & 2.92e+09\\
&& COSMOS & 6.65e+08 & 2.90e+08 & 1.68e+09 & 1.21e+08 & 2.76e+09\\\cline{1-8}
\cline{1-7}
\end{tabular}
\caption{Total kinetic energy of particles reaching the ground, for iron primary. Averages of 50 simulations are listed.}
\end{center}
\end{table}

\clearpage

\begin{table}
\begin{center}
\begin{tabular}[t]{c||c| c| ccccc}\hline
\multicolumn{8}{c}{Numbers of particles at the ground for proton primary}\\\hline
$\theta$&$\log_{10}E_0$ (eV) & simulation & photons & electrons & muons & hadrons & all \\\hline\hline
&18.5 & CORSIKA & 1.12e+10 & 1.72e+09 & 1.89e+07 & 8.50e+06 & 1.30e+10\\
&& COSMOS & 1.10e+10 & 1.74e+09 & 1.94e+07 & 8.83e+06 & 1.27e+10\\\cline{2-8}
0\ensuremath{^\circ}&19 & CORSIKA & 3.67e+10 & 5.69e+09 & 5.47e+07 & 2.56e+07 & 4.25e+10\\ 
&& COSMOS & 3.56e+10 & 5.72e+09 & 5.45e+07 & 2.20e+07 & 4.14e+10\\\cline{2-8}
&19.5 & CORSIKA & 1.17e+11 & 1.83e+10 & 1.61e+08 & 7.70e+07 & 1.35e+11\\ 
&& COSMOS & 1.15e+11 & 1.86e+10 & 1.53e+08 & 7.92e+07 & 1.34e+11\\\cline{1-8}
&18.5 & CORSIKA & 2.97e+09 & 3.71e+08 & 1.29e+07 & 3.34e+06 & 3.35e+09\\
&& COSMOS & 2.69e+09 & 3.60e+08 & 1.41e+07 & 4.19e+06 & 3.06e+09\\\cline{2-8} 
45\ensuremath{^\circ}&19 & CORSIKA & 1.14e+10 & 1.49e+09 & 3.92e+07 & 1.08e+07 & 1.29e+10\\
&& COSMOS & 1.04e+10 & 1.40e+09 & 3.99e+07 & 1.23e+07 & 1.18e+10\\\cline{2-8}
&19.5 & CORSIKA & 4.20e+10 & 5.57e+09 & 1.19e+08 & 3.48e+07 & 4.77e+10\\
&& COSMOS & 3.21e+10 & 4.35e+09 & 1.11e+08 & 3.56e+07 & 3.66e+10\\\cline{2-8}
\cline{1-7}
\end{tabular}
\caption{Total number of particles reaching the ground, for proton primary. Averages of 50 simulations are listed.}
\end{center}
\end{table}

\begin{table}
\begin{center}
\begin{tabular}[t]{c||c| c|ccccc}\hline
\multicolumn{8}{c}{Numbers of particles at the ground for iron primary}\\\hline
$\theta$&$\log_{10}E_0$ (eV) & simulation & photons & electrons & muons & hadrons & all \\\hline\hline
&18.5 & CORSIKA & 9.33e+09 & 1.37e+09 & 2.54e+07 & 9.96e+06 & 1.07e+10\\
&& COSMOS & 9.06e+09 & 1.38e+09 & 2.39e+07 & 1.01e+07 & 1.05e+10\\\cline{2-8}
0\ensuremath{^\circ}&19 & CORSIKA & 3.15e+10 & 4.68e+09 & 7.28e+07 & 2.97e+07 & 3.63e+10\\ 
&& COSMOS & 3.19e+10 & 4.95e+09 & 6.92e+07 & 3.19e+07 & 3.70e+10\\\cline{2-8}
&19.5 & CORSIKA & 1.07e+11 & 1.60e+10 & 2.10e+08 & 8.97e+07 & 1.23e+11\\ 
&& COSMOS & 1.02e+11 & 1.59e+10 & 1.96e+08 & 9.02e+07 & 1.19e+11\\\cline{1-8}
&18.5 & CORSIKA & 1.86e+09 & 2.35e+08 & 1.71e+07 & 3.60e+06 & 2.12e+09\\
&& COSMOS & 1.65e+09 & 2.17e+08 & 1.67e+07 & 3.96e+06 & 1.88e+09\\\cline{2-8} 
45\ensuremath{^\circ}&19 & CORSIKA & 6.95e+09 & 8.81e+08 & 5.01e+07 & 1.13e+07 & 7.89e+09\\
&& COSMOS & 6.42e+09 & 8.51e+08 & 4.85e+07 & 1.21e+07 & 7.33e+09\\\cline{2-8}
&19.5 & CORSIKA & 2.54e+10 & 3.24e+09 & 1.45e+08 & 3.46e+07 & 2.89e+10\\
&& COSMOS & 2.12e+10 & 2.80e+09 & 1.38e+08 & 3.60e+07 & 2.41e+10\\\cline{1-8}
\cline{1-7}
\end{tabular}
\caption{Total number of particles reaching the ground, for iron primary. Averages of 50 simulations are listed.}
\end{center}
\end{table}

\clearpage

\begin{table}
\begin{center}
\begin{tabular}[t]{c||c|ccccccccc}\hline
\multicolumn{9}{c}{Energy deposited to the air}\\\hline
primary& $\log_{10}E_0$ (eV) &18.5&18.75&19&19.25&19.5&19.75&20\\\hline\hline
&CORSIKA $\left<E_{\rm air}\right>/E_{0}$ & 0.69 & 0.69 & 0.70 & 0.70 & 0.70 & 0.70 & 0.71\\
proton&CORSIKA $\sigma_{E_{\rm air}}$/$\left<E_{\rm air}\right>$ & 0.015 & 0.013 & 0.014 & 0.014 & 0.013 & 0.011 & 0.010 \\\cline{2-9}
&COSMOS $\left<E_{\rm air}\right>/E_{0}$& 0.80 & 0.81 & 0.81 & 0.81 & 0.81 & 0.82 & 0.82 \\
&COSMOS $\sigma_{E_{\rm air}}$/$\left<E_{\rm air}\right>$ & 0.011 & 0.011 & 0.012 & 0.014 & 0.0094 & 0.011 & 0.011 \\\hline\hline
&CORSIKA $\left<E_{\rm air}\right>/E_{0}$& 0.66 & 0.67 & 0.68 & 0.68 & 0.68 & 0.69 & 0.69 \\
iron&CORSIKA $\sigma_{E_{\rm air}}$/$\left<E_{\rm air}\right>$ & 0.0074 & 0.0080 & 0.0078 & 0.0083 & 0.0079 & 0.0082 & 0.0071 \\\cline{2-9}
&COSMOS $\left<E_{\rm air}\right>/E_{0}$& 0.77 & 0.77 & 0.78 & 0.78 & 0.79 & 0.79 & 0.78 \\
& COSMOS $\sigma_{E_{\rm air}}$/$\left<E_{\rm air}\right>$ & 0.011 & 0.011 & 0.0098 & 0.010 & 0.0087 & 0.010 & 0.010\\\hline

\end{tabular}
\caption{Average of fraction and relative standard deviation of $E_{\rm air}$, the energy deposited to the air by the particles with energy above the threshold energy, in EASs of proton and iron primaries at the ground for $\theta$ = 70\ensuremath{^\circ}, which corresponds to $x_s = (875 / \cos 70\ensuremath{^\circ})$ g/cm$^{2}$. The average and standard deviation were calculated for 50 simulations for each set of parameters.}
\end{center}
\end{table}

\end{document}